\documentclass{aa}
\usepackage{psfig}

\def\,{\thinspace}

\font\sc=cmr10

\begin{document}

\title{CO and Dust in PSS~2322+1944 at a redshift of 4.12} 

\author{P. Cox\inst{1}
	\and A. Omont\inst{2}
	\and S.G. Djorgovski\inst{3}
	\and F. Bertoldi\inst{4} 
	\and J. Pety\inst{5}
        \and C.L. Carilli\inst{6}
 	\and \\ K.G. Isaak\inst{7} 
	\and A. Beelen\inst{1}
	\and R.G. McMahon\inst{8}
	\and S. Castro\inst{3,9}
	}
\offprints{P.~ Cox, pierre.cox@ias.u-psud.fr}

\institute{ Institut d'Astrophysique Spatiale, Universit\'e de Paris XI, 91405 Orsay, France
      \and  Institut d'Astrophysique de Paris, CNRS, 98bis boulevard Arago, F-75014 Paris, France
      \and  Astronomy Department, California Institute of Technology, Pasadena, CA 91125
      \and  Max-Planck-Institut f\"ur Radioastronomie, Auf dem H\"ugel 69, D-53121 Bonn, Germany
      \and  IRAM, 300 rue de la Piscine, F-38406 St-Martin-d'H\`eres, France
      \and  National Radio Astronomy Observatory, P.O. Box O, Socorro, NM 87801, USA
      \and  Cavendish Laboratory, Madingley Road, Cambridge CB3 0HE, UK
      \and  Institute of Astronomy, Madingley Road, Cambridge CB3 0HA, UK
      \and  Infrared Processing and Analysis Center 100-22, Caltech, Pasadena, CA 91125}

\date{Received 11 January 2002 / Accepted 8 March 2002}

\titlerunning{CO and Dust in  PSS~2322+1944 at $z = 4.12$}
\authorrunning{P. Cox et al.}

\abstract{  
  Using the IRAM interferometer we have detected J=4$\rightarrow$3 and
  5$\rightarrow$4 CO line emission toward the radio quiet quasar
  PSS~2322+1944. At a redshift of $z_{\rm CO}=4.1199$ this is the
  fourth and strongest detection of CO at $z>4$.  The
  velocity-integrated CO J=4$\rightarrow$3 and J=5$\rightarrow$4 line
  fluxes are $4.21 \pm 0.40$ and $3.74 \pm 0.56 \rm \, Jy \, km \,
  s^{-1}$, and the linewidth is $\approx 300 \, \rm km s^{-1}$.  The
  CO J=10$\rightarrow$9 was searched for but not detected with an
  upper intensity limit of 30~mJy.  The 1.35~mm (250~$\rm \mu m$ rest
  wavelength) continuum flux density is 7.5$\pm$1.3~mJy, in agreement
  with previous bolometer measurements at 1.2~mm with the 30-m IRAM
  telescope. The 3~mm (580~$\rm \mu m$ rest wavelength) continuum is
  not detected with a 3~$\sigma$ upper limit of 0.7~mJy.  We also
  report observations of the 450~$\rm \mu m$ continuum in
  PSS~2322+1944 using the SCUBA array at the JCMT. The quasar was
  detected with a 450~$\rm \mu m$ flux density of $\rm 79 \pm 19 \,
  mJy$. At the angular resolution of $4\farcs8 \times 2\farcs1$ at
  1.3~mm and $6\farcs2 \times 4\farcs9$ at 3.2~mm, the interferometer
  observations do not show evidence of spatial extension in the
  continuum or CO line emission. Assuming no gravitational 
  magnification, we estimate a molecular gas mass of 
  $\approx 2.5 \times 10^{11} \, \rm M_\odot$. The molecular gas is 
  warm ($\rm 40 - 100 \, K$) and dense ($\rm 10^{3.5} - 10^{4.1} \, cm^{-3}$). 
  The infrared-to-CO luminosity ratio is $\approx 185 \, \rm L_\odot 
  \, (K \, km \, s^{-1} \, pc^2)^{-1}$, comparable to the values found
  for ultraluminous infrared galaxies. The detection of CO emission 
  in this high redshift QSO provides further evidence that the radio 
  emission and the millimeter to submillimeter continuum emission are 
  predominantly powered by a starburst which is coeval with the AGN 
  activity.
\keywords{Galaxies: formation -- Quasars: emission lines --
Quasars: individual: PSS~2322+1944 -- Cosmology: observations 
-- Cosmology: early Universe -- Radio lines: galaxies}
}

\maketitle

\sloppy

\section{Introduction}

The study of dust and molecular gas in sources at high redshift has
opened up new ways to probe the physical conditions during the early
evolution of galaxies and to study the star formation history in the
early universe.  Spatially resolved observations of the molecular gas
content in the host galaxies of high-$z$ quasars are a key for
understanding the relationship between black hole formation and
spheroidal galaxy formation.  At redshifts $0.04 < z < 0.27$, the 
CO J=1$\rightarrow$0 line has been detected in 14 QSOs 
(Evans et al. 2001; Casoli \& Loinard 2002) indicating
large masses of molecular gas (a few $\rm 10^9 \, M_\odot$) 
fueling both the AGN and star formation in the QSO host galaxies. 
At higher redshifts,  although the lines are broad and weak 
and precise redshifts are difficult to predict, in the last decade 
$^{12}$CO emission was detected in 15 galaxies at $1.44 < z < 4.69$
(see Guilloteau 2002 and reference therein; Andreani et al. 2000; 
Barvainis et al. 2002). Most of the detections were obtained for 
high rotational transitions of CO (J=3$\rightarrow$2 to 9$\rightarrow$8) 
lines which are redshifted to millimeter wavelengths.  In a few cases, 
the J=2$\rightarrow$1 and 1$\rightarrow$0 lines were detected at
centimeter wavelengths (Carilli et al. 1999, 2002; Papadopoulos et
al. 2001). These CO observations imply that the high redshift galaxies
contain large amounts, a few $10^{10} \, \rm M_\odot$, of molecular
gas which are predominantly excited by giant starbursts with star
formation rates $\approx 10^3 \rm M_{\odot} yr^{-1}$ (see, e.g.,
Guilloteau et al.  1997, 1999).

In recent years we have concentrated our search for high-$z$ CO
emission to QSOs with strong (sub)millimeter dust continuum emission.
Based on the first surveys of (sub)millimeter emission of high-$z$
QSOs (McMahon et al. 1994; Omont et al. 1996a), we reported three
detections of CO emission in $z>4$ sources: BR~1202$-$0725 (Omont et
al. 1996b), BRI~1335$-$0415 (Guilloteau et al. 1997) and
BRI~0952$-$0115 (Guilloteau et al. 1999).  We recently extended these
surveys for continuum emission using the Max-Planck millimeter
bolometer (MAMBO) array at the 30-meter (Omont et al. 2001; Carilli et
al. 2001a) and SCUBA at the JCMT (Isaak et al. 2002).  Continuum
emission was detected toward 35 new high-$z$ QSOs: Omont et al. (2001)
and Isaak et al. (2002) observed a total of 75 bright quasars at $z >
3.8$ from the Palomar Sky Survey (PSS) sample and detected 19 sources;
Carilli et al. (2001a) made a deeper survey of a sample of 41 QSOs at
$z > 3.7$ selected from the Sloan Digital Sky Survey (SDSS) and
detected 16 of them.

\begin{figure}[bh!]
\centerline{\psfig{figure={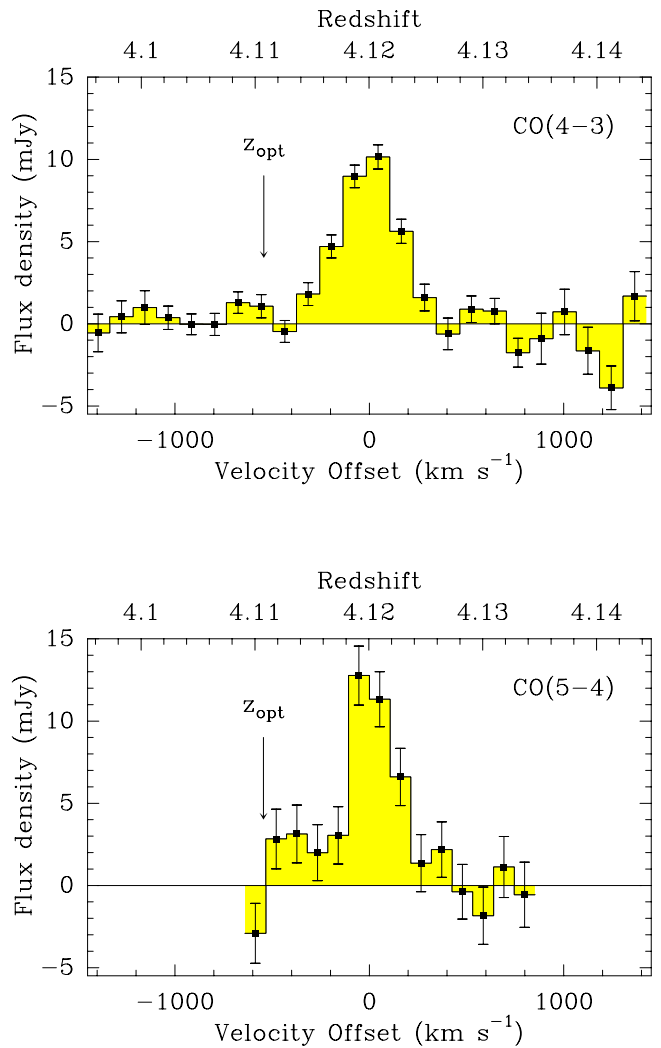},width=8.5cm}}
	\caption{Spectra of the CO J=4$\rightarrow$3 and
	J=5$\rightarrow$4 lines toward PSS~2322+1944.  Error bars are
	$\pm 1\sigma$. The velocity scale corresponds to frequencies
	of 90.05 and 112.55~GHz, or a redshift $z_{\rm CO} = 4.1199
	\pm 0.0008$. The redshift derived from multiple, associated
	narrow absorption lines in a Keck optical spectrum is
	$z_{\rm opt} = 4.11075$, which is indicated by an arrow. The
	weak continuum was not subtracted.}
\label{figure1}
\end{figure}

PSS~2322+1944, discovered by Djorgovski et al. (in prep.), is the
brightest QSO in the survey of $z>3.8$ PSS quasars by Omont et
al. (2001) and Isaak et al. (2002), with flux densities of $9.6\pm
0.5$ and $22.5\pm 2.5$~mJy at 250 and 350~GHz, respectively. The
luminosity and implied dust mass are estimated\footnote{for a
$\Lambda$ cosmology ($H_0=65\rm~km~s^{-1}~Mpc^{-1}$,
$\Omega_\Lambda=0.7$, $\Omega_m=0.3$) adopted throughout this paper,
the luminosity distance $D_L$ is $\rm 3.99 \times 10^4 \, Mpc$.}  to
be $\rm \sim 2.7 \times 10^{13} \, L_{\odot}$ and $\rm \sim 1.6 \times
10^9 \, M_{\odot}$ (see Sect.~3). Under the assumption that the far-infrared
luminosity arises exclusively from young stars, the implied star
formation rate is a few $\rm 10^3 \, M_{\odot} yr^{-1}$.
PSS~2322+1944 was also detected with the VLA at 1.4~GHz with a total
flux density of $98\pm20 \, \rm \mu Jy$ being spatially extended on a
scale of 1.5$^{\prime\prime}$ (Carilli et al.  2001b).  The ratio of
radio to millimeter flux agrees with the radio-to-far infrared
correlation for star-forming galaxies.  PSS~2322+1944 is also an
exceptional optical source as described by Djorgovski et al. (in prep.).
From a Keck optical spectrum a redshift was derived from multiple,
associated narrow absorption lines, yielding $z_{\rm opt} = 4.11075 \pm
0.0005$.

\begin{figure*}[th!]
\vspace{1.8cm}
\centerline{\psfig{figure={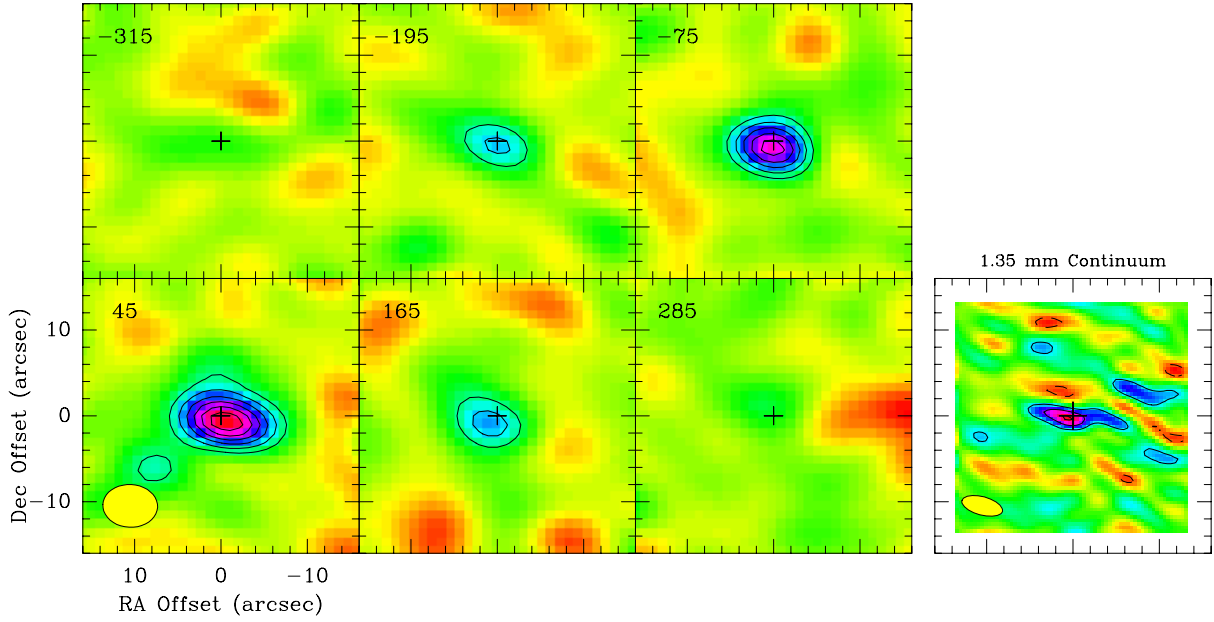},width=16cm,angle=-90}}
\vspace{-1cm}
	\caption{{\it (a)} Channel maps of the CO J=4$\rightarrow$3 line
	toward PSS~2322+1944.  The channel width is 35~MHz, i.e., $\rm
	120 \, km \, s^{-1}$. The contour step is 1.33~mJy
	beam$^{-1}$.  The offset velocity is indicated in the upper
	left corner of each map. The zero velocity corresponds to
	90.05~GHz, i.e., $z=4.1199$.  The beam of $\rm 6.25 \times
	4.98 \, arcsec^2$ is shown in the lower left panel.  {\it (b)}
	The lower right panel shows the 1.35~mm continuum map.
	Contour steps are 2~mJy beam$^{-1}$, and the r.m.s. noise
	1.5~mJy beam$^{-1}$. The beam of $\rm 4.85 \times 2.14 \,
	arcsec^2$ is shown in the lower left corner.  Offsets are
	relative to the optical (and radio) position (indicated by a
	cross), R.A. 23:22:07.25, Dec. 19:44:22.08 (J2000.0). }
\label{figure2}
\end{figure*}

\section{Observations and Results}

Observations were made with the IRAM Plateau de Bure interferometer
during 5 nights in June 2001.  We used the standard CD configuration
(4 antennas) which results in a beam of $4\farcs8 \times 2\farcs1$ at
1.3~mm and $6\farcs2 \times 4\farcs9$ at 3.2~mm. Dual frequency
receivers were used to search simultaneously for CO J=4$\rightarrow$3
and dust emission at 3 and 1.35~mm, and in a second series of
observations for CO J=5$\rightarrow$4 and 10$\rightarrow$9 emission.
During the first observations, the receivers were tuned at 90.03
and 90.06~GHz to search for the J=4$\rightarrow$3 emission line.  
A subsequent series of observations was done with the 3~mm receiver
centered at 112.6~GHz (at the position of the redshifted CO
J=5$\rightarrow$4) and the 1.3~mm receiver tuned at 225.0~GHz to
search for CO J=10$\rightarrow$9. The 3~mm receivers were tuned in
single sideband and the 1.3~mm receivers in double sideband. Typical
SSB system temperatures were $\rm \approx 150 \, K$ at 3~mm and $\rm
\approx 400 \, K$ at 1.3~mm. Amplitude and phase calibration were done
using 3C~454.3 (6.1, 5.3~Jy) and 2230+114 (2.1, 1.2~Jy) where the 
number in round brackets are the flux densities at 3 and 1.3~mm, 
respectively, at the date of the observations. Phase noise was 
stable on all baselines both at 3~mm (rms 5$\rm ^o$--17$\rm ^o$) 
and at 1.3~mm (rms 13$\rm ^o$--45$\rm ^o$). During the observations, 
the water vapour varied between 3 and 6~mm and the seeing conditions 
varied between $0\farcs8$ and $1\farcs3$, i.e. much smaller than the 
synthesized beams at 3~mm and 1~mm. The total integration time was 
about 18~hours for the CO J=4$\rightarrow$3 line, and 12~hours for 
the CO J=5$\rightarrow$4, the CO=10$\rightarrow$9 and the 1.35~mm 
continuum. The data were reduced, calibrated and analyzed using the 
standard IRAM programs C{\small LIC} and M{\small APPING}.  The final 
spectra are shown in Fig.~1 and, for the CO J=4$\rightarrow$3 emission, 
Fig.~2a also shows the velocity channel maps.

PSS~2322+1944 was detected in the continuum at 1.35~mm with a flux
density of 7.5$\pm$1.3~mJy, as derived from the image sideband
measurements. This value is consistent with the flux density measured
at 1.20~mm with MAMBO at the 30-meter ($9.6\pm0.5 \rm \, mJy$) when
the spectral index of dust emission is taken into account. Within the
astrometric uncertainties ($\pm 0\farcs3$), the continuum source
corresponds is to the optical (and radio) position, RA 23:22:07.25, 
Dec 19:44:22.08 (J2000.0) from Carilli et al. (2001b). At the angular 
resolution of our observations, the source is not resolved (Fig.~2b).  
PSS~2322+1944 is not detected in the continuum at 3~mm with a flux 
density of $0.40 \pm 0.25 \rm \, mJy$ as derived from the line-free 
channels. Using a dust spectral index of 3.5, this is consistent 
with the 3\,mm continuum value of $\approx 0.3$\,mJy expected from 
the 1.35~mm flux density (Fig.~3).

PSS~2322+1944 was also observed with the wide-band 450~$\rm \mu m$ and
850~$\rm \mu m$ filters on SCUBA at the JCMT in December 2001 under
good and stable weather conditions with $\tau_{\rm 850 \, \mu m} \sim
0.06$. The observations, calibration and data reduction were done as
explained in Isaak et al. (2002). The source was detected in the
continuum at 450~$\rm \mu m$ (corresponding to a rest wavelength of
88~$\rm \mu m$) with a flux density of 75$\pm$19~mJy. At 850~$\rm \mu
m$, the flux density of PSS~2322+1944 is 24$\pm$2~mJy, which is
entirely consistent with the 22.5$\pm$2.5~mJy value reported in Isaak
et al. (2002). Fig.~3 presents the spectral energy distribution of
PSS~2322+1944 with all the currently available photometric data.

Both the CO J=4$\rightarrow$3 ($\nu_{\rm rest} = 461.0408~ \rm GHz$)
and J=5$\rightarrow$4 ($\nu_{\rm rest} = 576.2679~ \rm GHz$) are
clearly detected toward the position of the continuum emission. The
J=4$\rightarrow$3 and 5$\rightarrow$4 lines are found at frequencies
of 90.05 and 112.55~GHz (Table~1) corresponding to a redshift of
$z_{\rm CO} = 4.1199 \pm 0.0008$, close to the redshift derived from
the optical spectrum ($z_{\rm opt} = 4.11075$). The difference
corresponds to a velocity difference of $530 \, \rm km \, s^{-1}$
(Fig.~1).  The CO J=10$\rightarrow$9 emission line was not detected
with a $3 \sigma$ upper intensity limit of 30~mJy (Table~1).  The
integrated line fluxes of the CO J=4$\rightarrow$3 and
J=5$\rightarrow$4 lines are $\rm 4.24 \pm 0.33 \, Jy \, km \, s^{-1}$
and $\rm 3.74 \pm 0.56$, respectively.  PSS~2322+1944 is the strongest
CO emitter at high $z$, even stronger than the lensed quasar
APM~08279+5255 (Downes et al. 1999) -- see Table~2.  From Gaussian
fits, the line widths are found to be $\rm \approx 300 \, km \,
s^{-1}$ (Table~1) which is comparable to the width in the three $z>4$
CO sources detected so far (Table~2).

The channel maps in the CO J=4$\rightarrow$3 emission line (Fig.~2a)
(as well as the J=5$\rightarrow$4 channel maps not shown) do not show
evidence for extension and/or position shifts with velocity.  The
3.2~mm CO data are limited by the angular resolution of the
observations ($6\farcs2 \times 4\farcs9$) and other array
configurations will be needed to probe further whether the emission of
PSS~2322+1944 is extended.


\begin{table*}[]
\caption{Observed Properties of CO Lines in PSS~2322+1944}
\begin{center}
\begin{tabular}{cccccccc}
\hline
\hline 
Line &  $\rm \nu_{\rm obs}$ & Peak Int.  &  $\Delta V_{\rm FWHM}$   &  $I_{\rm CO}$  & Continuum  &  $L^\prime_{\rm CO}$ & $L_{\rm CO}$  \\

     &    [GHz] &  [mJy] & [$\rm km \, s^{-1}$] &  [Jy~km~s$^{-1}$] & [mJy] & [$\rm 10^{11} \, K \, km \, s^{-1} \, pc^2$] & [$10^8 \, L_\odot$] \\ 

\hline \\[-0.2cm]

CO(4$\rightarrow$3)  &    90.05  & 10.5    &  375$\pm$41 &   4.21$\pm$0.40   &  0.40$\pm$0.25 & 2.0 &  6.3 \\      

CO(5$\rightarrow$4)  &   112.55  & 12.0    &  273$\pm$50 &   3.74$\pm$0.56   &    --          & 1.1  & 7.0 \\

CO(10$\rightarrow$9) &   225.00  & $< 30$  &     --      &    $<$5.2$\dagger$        &  $7.5 \pm 1.3$ & $< 0.4$ & $< 19$ \\       

\hline
\end{tabular}
\end{center}
NOTE. -- $^\dagger$ Adopting a line width of $\rm 300 \, km \, s^{-1}$
\end{table*}


\begin{table*}[]
\caption{Comparison of the CO Results of PSS~2322+1944 with other $z > 3.5$ Sources}
\begin{center}
\begin{tabular}{lcccccc}
\hline
\hline 
Source       & $z$  & \multicolumn{2}{c}{CO(4$\rightarrow$3)} &  \multicolumn{2}{c}{CO(5$\rightarrow$4)} &  Ref.  \\

             &      &   $I_{\rm CO}$   &  $\Delta V_{\rm FWHM}$  &
             $I_{\rm CO}$ & $\Delta V_{\rm FWHM}$ & \\

             &      & [Jy~km~s$^{-1}$] & [km~s$^{-1}$] & [Jy~km~s$^{-1}$] & [km~s$^{-1}$] &        \\     

\hline \\[-0.2cm]

PSS2322+1944    & 4.12 & 4.21$\pm$0.40 & 375$\pm$41 &  3.74$\pm$0.56 & 273$\pm$50 &   [1]       \\

BRI~1335$-$0415 & 4.41 &               &            &  2.80$\pm$0.30 & 420$\pm$60 &   [2]  \\

BRI~0952$-$0115 & 4.43 &               &            &  0.91$\pm$0.11 & 230$\pm$30 &   [3]  \\

BR~~1202$-$0725 & 4.69 & 1.50$\pm$0.30 & 280$\pm$30 &  2.40$\pm$0.30 & 320$\pm$35 &   [4,5]  \\

APM~08279+5255  & 3.87 & 3.70$\pm$0.50 & 480$\pm$35 &                &            &   [6]  \\

4C~60.07        & 3.79 & 2.50$\pm$0.43 & $\ge 1000$ &                &            &   [7]  \\

6C~1909+722     & 3.53 & 1.62$\pm$0.30 & 530$\pm$70 &                &            &   [7]  \\

\hline
\end{tabular}
\end{center}
NOTE. -- [1] This paper [2] Guilloteau et al. (1997) 
[3] Guilloteau et al. (1999)  [4] Ohta et al. 1996 
[5] Omont et al. (1996b) [6] Downes et al. (1999) 
[7] Papadopoulos et al. (2000) 
\end{table*}

\section{Discussion}

The observed CO J=4$\rightarrow$3 and 5$\rightarrow$4 line fluxes
imply intrinsic CO line luminosities, $L^\prime_{\rm CO} = 2.0$ and
$\rm 1.1 \times 10^{11} \, K \, km \, s^{-1} \, pc^2$, respectively,
or expressed in solar luminosities, $L^\prime_{\rm CO}= 6.3$ and $7.0
\times 10^8 \, \rm L_\odot$ (Table~1). $L^\prime_{\rm CO}$ is
proportional to line brightness (Rayleigh-Jeans) temperature
integrated over the area of the source: the ratio of luminosities in
two CO transitions originating from the same area is a measure of the
line brightness ratio and therefore an indicator of the physical
conditions in the molecular gas (see Solomon et al. [1997] for 
a discussion of $L^\prime_{\rm CO}$ and $L_{\rm CO}$).

From the $L^\prime_{\rm CO}$ we derive brightness (Rayleigh-Jeans)
temperature ratios
$T_b$[CO(4$\rightarrow$3)]/$T_b$[CO(5$\rightarrow$4)]~$ = 1.76 \pm
0.25$ and
$T_b$[CO(10$\rightarrow$9)]/$T_b$[CO(5$\rightarrow$4)]$<$0.34.  The
former ratio is comparable to the values of 1.38 derived for the
$z=2.56$ quasar H~1413 (the Cloverleaf) - see Barvainis et al. (1997) - 
and of 1.02 found for the $z=4.69$ quasar BR~1202$-$0725 (Omont et al. 1996a).
These ratios indicate that the molecular gas in PSS~2322+1944 must be
warm and dense. Using the results of a Large Velocity Gradient model
for high $z$ galaxies (Sakamoto 1999), we used the brightness
temperature ratios to constrain the molecular gas density in
PSS~2322+1944.  For a gas kinetic temperature of $\rm \approx 40-100
\, K$, in accordance with the temperature derived from the thermal
dust emission spectrum (see below), and adopting a value of $\rm
10^{-6} \, pc/(km \, s^{-1})$ for the CO abundance per unit velocity
gradient of the molecular gas ($X({\rm CO})/dv/dr$), the
$T_b$[CO(4$\rightarrow$3)]/$T_b$[CO(5$\rightarrow$4)] ratio constrains
the density of the molecular gas to be of the order of $\rm 10^{3.5} -
10^{4.1} \, cm^{-3}$. These values for the gas temperature and density
are also consistent with the line brightness ratio
$T_b$[CO(4$\rightarrow$3)]/$T_b$[CO(1$\rightarrow$0)] $\approx 1.4$
derived from the CO(1$\rightarrow$0) line flux measured with the VLA
in PSS~2322+1044 (Carilli et al. [in prep.] - a more detailed analysis
of the excitation conditions will be given in this paper). The gas
density in PSS~2322+1944 is comparable to the high densities ($n(\rm
H_2) \approx 10^4 \, cm^{-3}$) found in the highly excited gas in the
nuclear region of the starburst galaxy M~82 (G\"usten et al. 1993; Mao
et al. 2000) and in nearby ultraluminous infrared galaxies (e.g.,
Solomon et al. [1992]).

Assuming a standard Galactic value of $\rm 4.6 \, M_\odot \, (K \, km
\, s^{-1} \, pc^2)^{-1}$ for the conversion factor of molecular mass
to $L^\prime _{\rm CO(1-0)}$, and using the above line brightness
ratio $T_b$[CO(4$\rightarrow$3)]/$T_b$[CO(1$\rightarrow$0)], we infer
a molecular gas mass of $\rm \approx 6 \times 10^{11} \, M_\odot$.  
Based on a study of ultraluminous infrared galaxies (ULIRGs), Solomon
et al. (1997) showed that the molecular gas mass to CO luminosity
conversion factor is likely to be lower (by a factor $\approx 3$) in
the extreme environments of luminous infrared galaxies where the
molecular clouds are expected to be hotter and denser than in the
Galaxy (see also Combes et al. 1999). In the case of Arp~220, Scoville
et al. (1997) derived a conversion factor $\approx 0.45$ times the
Galactic value.  The above gas mass is therefore likely to be
overestimated by a factor 2 to 3. In the following we will adopt a  
conversion factor of  $\rm 1.8 \, M_\odot \, (K \, km \, s^{-1} \, pc^2)^{-1}$
(i.e., a factor of 2.5 times smaller than the Galactic conversion
factor), which translates into a molecular gas mass of 
$M_{\rm H_2} \approx 2.5 \times 10^{11} \rm M_\odot$ for PSS~2322+1944.

Compared with the dust mass $M_{\rm dust} \approx 1.6 \times 10^9 \, \rm
M_\odot$ derived from the 1.35~mm continuum flux density (see Omont et
al. 2001), the implied gas to dust mass ratio for PSS~2322+1944 
is $M({\rm H_2})/M_{\rm dust} \approx 150$, comparable to the ratios 
derived in bright nearby galaxies (Dunne et al. 2000), ULIRGs 
(Solomon et al. 1997), or other high redshift CO sources 
(Guilloteau et al. 1999).  This similarity is also reflected 
in the $L_{\rm FIR}/L^\prime_{\rm CO(1-0)}$ ratios of these sources. 
In the case of PSS~2322+1944, $L_{\rm FIR}/L^\prime_{\rm CO(1-0)} 
\approx 185 \, \rm L_\odot \, (K \, km \, s^{-1} \, pc^2)^{-1}$, 
close to the median value of 160 found for a sample of 37 ULIRGs 
by Solomon et al. (1997).

\begin{figure}[]
  \centerline{\psfig{figure={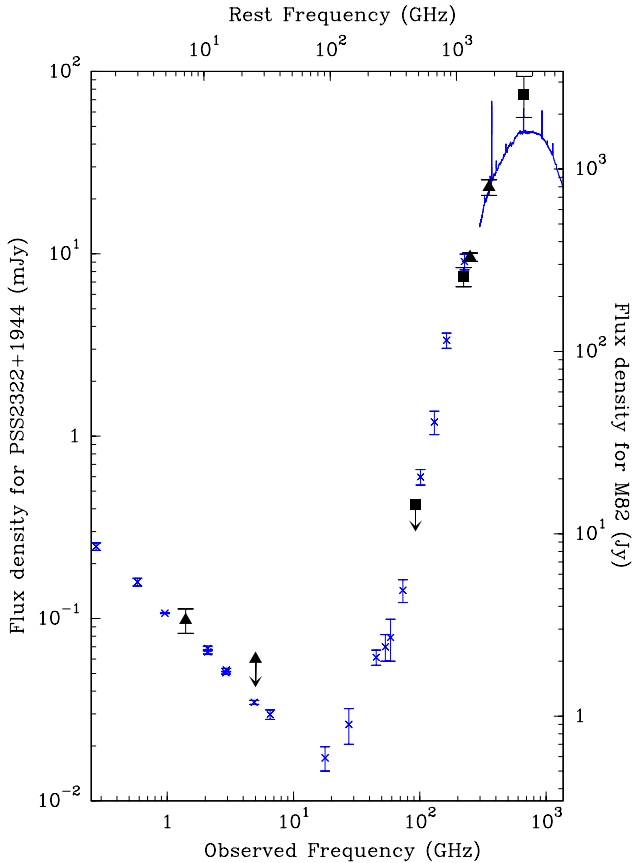},width=8.5cm}} 
  \caption{ Spectral energy distribution of PSS~2322+1944.  {\it
  	Triangles:} 850~$\rm \mu m$ (Isaak et al.  2001); 1.2~mm
  	(Omont et al.  2001); 1.4 and upper limit at 5~GHz (Carilli et
  	al. 2001b); {\it squares:} 450~$\rm \mu m$, 1.35~mm and the $3
  	\sigma$ upper limit at 3~mm (this paper).  For comparison, the
  	radio-to-infrared spectral energy distribution of the
  	starburst galaxy M~82 is shown, red-shifted to $z= 4.12$ and
  	normalized to the flux density of PSS~2322+1944 at the
  	observed wavelength of 850~$\rm \mu m$: the crosses show all
  	the currently available photometric data, and the continuous
  	line represents the ISO LWS spectrum (from Colbert et al.
  	1999). The left- and right-hand flux density scale are
  	adapted for PSS~2322+1944 and M~82, respectively.}
\label{figure3}
\end{figure}

Figure~\ref{figure3} presents the radio to infrared flux densities of
PSS~2322+1944 (including the new measurement at 450~$\rm \mu m$ and
1.35~mm) compared with the spectral energy distribution of M~82,
which we red-shifted to $z= 4.12$ and normalized to the flux density 
of PSS~2322+1944 at the observed wavelength of 850~$\rm \mu m$. 
Note that M~82 is about 100 times less luminous than PSS~2322+1944.  
The two spectral energy distributions are remarkably
similar in the far-infrared/radio, although the peak intensity of
PSS~2322+1944 is somewhat higher.  For a dust emissivity index $\beta
= 1.6$, the dust emission is well fitted with a temperature of $T_d
\approx 47 \, \rm K$, which is similar to the dust temperature
derived in M~82 under the same assumptions, i.e. $T_d \approx 45 \,
\rm K$.  This further supports the idea that in PSS~2322+1944 the
starburst dominates the dust heating and the radio emission (see also
Omont et al. [2001] and Carilli et al. [2001b]).  Further confirmation
may come from a comparison of their mid- and near-infrared spectra.
Similarities with the far-infrared and radio SED of M~82 were also 
noted for the QSOs BRI~1335--0417 at $z=4.4$ (Carilli et al. 1999) 
and BR~1202--0725 at $z=4.7$ (Kawabe et al. 1999; Yun et al. 2000).

Djorgovski et al. (in prep.) found that the quasar PSS~2322+1944 has a
close companion (separated by about $1^{\prime\prime}$) with the same spectrum 
within the measurement uncertainties.  It is not yet known if this is a 
case of a gravitational lensing, or a binary quasar.  In the former case, 
the luminosities and dust masses deduced here should be lowered by a
factor of a few.  In the latter case, we may be seeing an interaction
or a merger of two quasar hosts.

\section{Conclusions}

With an apparent CO luminosity greater than that of the strongly lensed
quasar APM~08279+5255, the $z = 4.12$ quasar PSS~2322+1944 is the
strongest CO emitter detected to date at redshifts larger than
3.5. Assuming no gravitational magnification, we estimate a molecular
gas mass of $\approx 2.5 \times 10^{11} \, \rm M_\odot$, and a far-infrared
luminosity of $\rm \approx 2.7 \times 10^{13} \, L_{\odot}$.  The
spectral energy distribution and large luminosity suggest that a
massive starburst takes place in PSS~2322+1944, which may be related
to the formation of the core of an elliptical galaxy.

The exceptional brightness of PSS~2322+1944 makes it a good target for
further observations of other CO transitions, in particular those from
lower levels which will constrain the physical conditions of the bulk
of the molecular gas. Higher spatial resolution measurements are also
needed to see whether the line or continuum emission is extended, as
it was seen for the radio and optical emission (Carilli et al. 2001b;
Djorgovski et al., in prep.).

The detection of CO in another strong (sub)millimeter continuum
high-$z$ quasar confirms a frequent correlation between the 3~mm CO
peak intensity and the 1.3~mm continuum flux, with a typical ratio of
about unity. This relation shows that systematic searches for CO
emission in strong thermal dust continuum quasars are promising with
current instrumentation, provided that the redshift is known with high
enough accuracy.  Systematic (sub)millimeter continuum surveys of high
redshift, radio-quiet quasars are therefore needed to find strong
continuum sources towards which CO emission can be searched. Such
studies promise to further our understanding of the physical and
chemical properties of the most energetic sources in the early
Universe.  Interferometric observations, especially with ALMA and
EVLA, will eventually be able to show the spatial distribution of the
molecular gas and its relation to the stars and ionized gas.

\acknowledgements 

R. Neri and M. Grewing are gratefully acknowledged for their support
of this program, and D. Downes for useful discussions. We also thank
R. Lucas for help with the data reduction and the referee, A.S. Evans,
for comments which improved the contents of this paper. The IRAM 
Plateau de Bure staff is kindly acknowledged for its efficient 
assistance.  IRAM is supported by INSU/CNRS (France), MPG (Germany), 
and IGN (Spain).  The JCMT is operated by JAC, Hilo, on behalf of 
the parent organisations of the Particle Physics and Astronomy 
Research Council in the UK, the National Research Council in 
Canada and the Scientific Research Organisation of the Netherlands. 
The National Radio Astronomy Observatory (NRAO) is a facility of 
the National Science Foundation, operated under cooperative 
agreement by Associated Universities, Inc. SGD and SMC acknowledge 
a partial support from the Bressler Foundation.


\end{document}